\documentclass[conf, onecolumn]{ao4elt3}
\usepackage{epsfig}
\usepackage{subfigure}
\usepackage{lscape}
\usepackage{graphicx}
\usepackage[varg]{txfonts} 
\usepackage[latin1]{inputenc}
\usepackage{deluxetable}

\def\procspie{Proc. SPIE}

\def\newblock{\hskip .11em plus .33em minus .07em}

\def\lesssim{\mathrel{\hbox{\rlap{\hbox{\lower4pt\hbox{$\sim$}}}\hbox{$<$}}}}
\def\gtrsim{\mathrel{\hbox{\rlap{\hbox{\lower4pt\hbox{$\sim$}}}\hbox{$>$}}}}

\def\arcsec{\hbox{$^{\prime\prime}$}}

\def\farcm{\hbox{.\kern -0.7ex\raisebox{.9ex}{\scriptsize$\prime$}}}
\def\farcs{\hbox{\kern 0.13ex.\kern -0.95ex%
\raisebox{.9ex}{\scriptsize$\prime\prime$}\kern -0.1ex}}

\def\apj{ApJ}

\def\aaps{A\&AS}

\def\it{}

\def\aap{AAP}

\def\pasp{PASP}

\def\lesssim{\mathrel{\hbox{\rlap{\hbox{\lower4pt\hbox{$\sim$}}}\hbox{$<$}}}}
\def\gtrsim{\mathrel{\hbox{\rlap{\hbox{\lower4pt\hbox{$\sim$}}}\hbox{$>$}}}}

\def\arcsec{\hbox{$^{\prime\prime}$}}

\def\farcm{\hbox{.\kern -0.7ex\raisebox{.9ex}{\scriptsize$\prime$}}}
\def\farcs{\hbox{\kern 0.13ex.\kern -0.95ex%
\raisebox{.9ex}{\scriptsize$\prime\prime$}\kern -0.1ex}}

\def\arcsec{''}

\def\la{\mathrel{\hbox{\rlap{\hbox{\lower4pt\hbox{$\sim$}}}\hbox{$<$}}}}
\def\ga{\mathrel{\hbox{\rlap{\hbox{\lower4pt\hbox{$\sim$}}}\hbox{$>$}}}}

\def\arcsec{\hbox{$^{\prime\prime}$}}

\def\farcm{\hbox{$.\mkern-4mu^\prime$}}
\def\farcs{\hbox{$.\!\!^{\prime\prime}$}}

\def\fnum@figure{{\rmfamily Fig.\space\thefigure.---}}%
\def\fnum@table{{\rmfamily Table \thetable:}}%
\def\fnum@plate{{\bfseries Plate \theplate.}}%
\def\fps@figure{bp}%
\def\fps@table{bp}%
\def\fps@plate{bp}%
\def\eps@scaling{1.0}%

\begin{document}
\pagestyle{plain}
\title{ASTROMETRY IN THE GALACTIC CENTER WITH THE THIRTY METER TELESCOPE}
\author{Sylvana Yelda\inst{1}\thanks{syelda@astro.ucla.edu}, Leo Meyer\inst{1}, Andrea Ghez\inst{1} \and Tuan Do\inst{2}} 
\institute{UCLA Department of Physics and Astronomy, Los Angeles, CA 90095-1547 USA \and Dunlap 
Institute for Astronomy and Astrophysics, University of Toronto, Toronto M5S 3H4, ON, Canada}

\abstract{
We report on the expected astrometric performance of the Thirty Meter Telescope's
InfraRed Imaging Spectrometer (IRIS) as determined using simulated images
of the Galactic center.  This region of the Galaxy harbors a supermassive
black hole and a dense nuclear stellar cluster, thus providing an ideal laboratory
for testing crowded-field astrometry with the IRIS imager.  Understanding the sources
of astrometric error is also important for making precision measurements 
of the short-period stars orbiting the supermassive black hole in order to probe the 
curvature of space-time as predicted by General Relativity. 
Various sources of error are investigated, including read-out and photon 
noise, spatially variable point spread functions, confusion, static distortion for the 
IRIS imager, and the quadratic probe arm distortion.  Optical distortion is the 
limiting source of error for bright stars ($K <$ 15), while fainter sources will be 
limited by the effects of source confusion. A detailed astrometric error budget
for the Galactic center science case is provided.
}

\maketitle

\section{Introduction}
\label{intro}
The InfraRed Imaging Spectrometer (IRIS; Larkin et al. \cite{larkin10}) is a 
first-light instrument for the
Thirty Meter Telescope (TMT) that will operate behind the Narrow-Field Infrared 
Adaptive Optics System (NFIRAOS; Herriot et al. \cite{herriot12}) and that will 
provide images with the highest resolution and Strehl 
ever taken in the infrared. The imaging camera will operate at wavelengths
$\lambda$ = 0.8 - 2.5 $\mu$m with a plate scale
of 4 mas/pixel and a field of view of 17$\farcs$2 $\times$ 17$\farcs$2. 
At $\lambda$ = 2 $\mu$m, the diffraction-limited resolution of 
$\Theta \sim$ 14 mas will be achieved, which will allow for unprecedented 
astrometric precision.

The astrometric performance of IRIS can be tested using simulations of the 
Galactic center (GC), which harbors a supermassive black hole (SMBH) and a dense 
nuclear stellar cluster. In this report, various 
sources of astrometric error are investigated for single-epoch observations of
the GC, and a detailed astrometric error budget 
for this science case is provided.  Future work will include
multi-epoch simulations in order to determine the precision with which 
post-Newtonian effects in the orbits of short-period stars can be measured.

\section{Methods}
\label{sec:methods}
A series of systematic tests are performed, each of which investigates the contributions
of individual error sources to the astrometric precision of stars in the Galactic center.  
The sources of error are incorporated in single-epoch simulated images, and include:
\begin{itemize}
\item photon, detector, and thermal noise,
\item On-Instrument Wave Front Sensor (OIWFS) probe arm positioning error,
\item static distortion,
\item the combination of probe arm and static distortion,
\item source confusion.
\end{itemize}

\subsection{Simulated Images of the Galactic Center}
\label{sec:images}
Ten PSF images of 20 seconds each were simulated by the TMT Adaptive Optics group.
These ten phase screen realizations 
of spatially-variable PSFs (a grid of PSFs constant over 2$\arcsec\times$2$\arcsec$) 
were computed specifically for a chosen guide star constellation at the GC with the
following offsets from Sgr A*: ($x$, $y$) = (15.3$\arcsec$,-10.9$\arcsec$), 
(-9.5$\arcsec$, 11.8$\arcsec$), and (-5.9$\arcsec$, -17.6$\arcsec$). 
The Strehl ratio of the simulated PSFs ranged from $\sim$70-80\%.
In addition, each PSF was computed for five specific near-infrared wavelengths 
(1.908, 2.067, 2.12, 2.173, 2.332 $\mu$m). The GC observations will most likely be 
carried out using a narrow-band filter in order to avoid saturation. However, 
selecting a single wavelength leads to a highly-structured and unrealistic 
halo dominated by diffraction effects (``satellite spots''). This is somewhat 
unrealistic since those features will get smeared out for three reasons: 
(1) they will rotate with galactic angle, (2) the wavelength interval 
is small but finite, and (3) an implementation error (not accounted for here) will lead to 
more diffuse flux in the halo. Thus, we have chosen to mimic these three effects by 
averaging over all five wavelengths and including noise properties of the narrowband
filter $K_{cont}$ ($\lambda$ = 2.27 $\mu$m, $\Delta\lambda$ = 0.03 $\mu$m). 
The background magnitude is estimated at $K_{cont}$ = 13.4 mag.
Each image is simulated using one of the ten phase screen realizations of
spatially-variable PSFs.  The resulting images are 4096$\times$4096 pixels with 
a plate scale of 4 mas/pixel. In some tests, we also simulate images
with a constant PSF over the field, in which case a single PSF is selected from
the available grid.

\begin{figure}[h]
\centering
\includegraphics[width=0.45\textwidth]{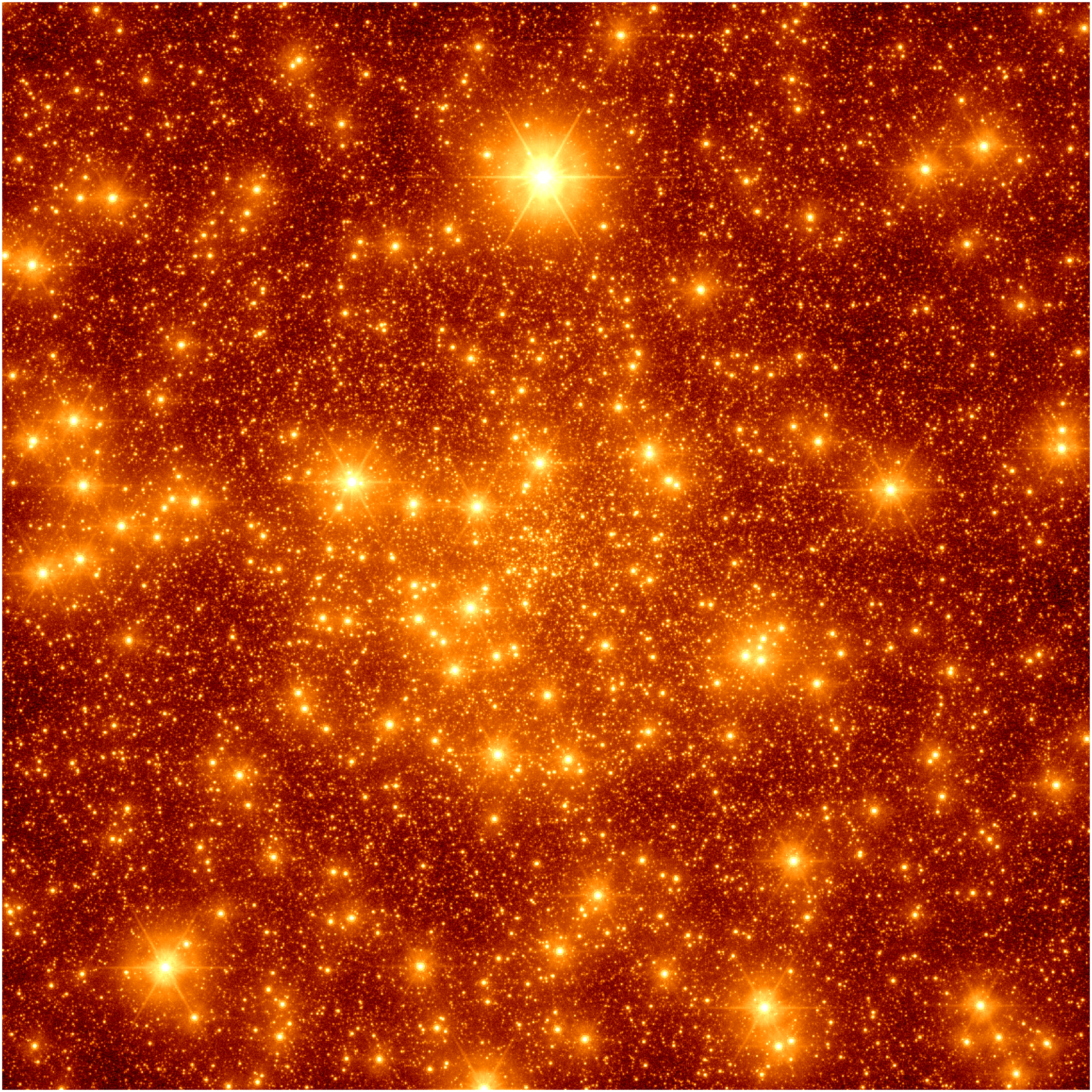}
\includegraphics[width=0.45\textwidth]{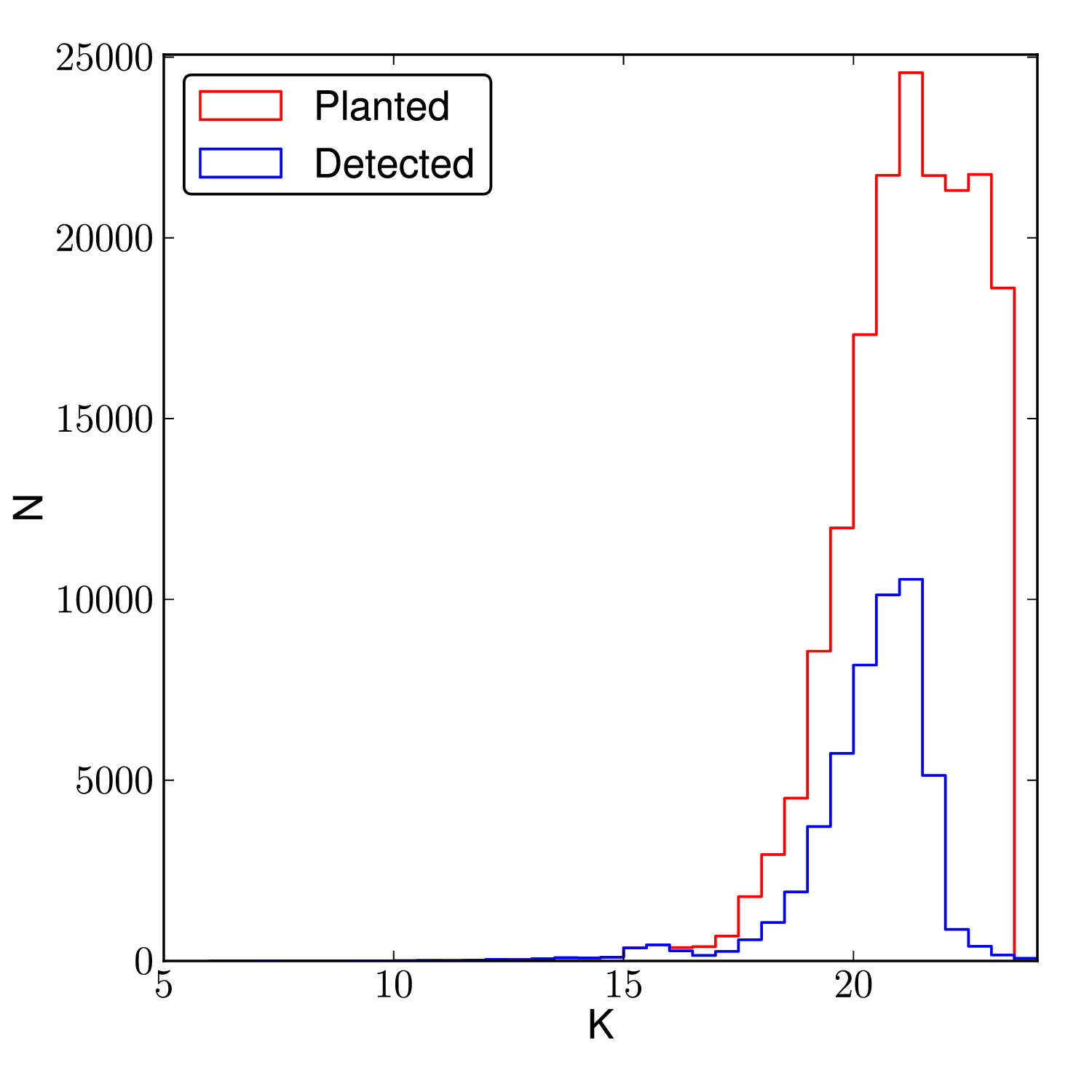}
\caption{{\em Left}: Simulated TMT/IRIS near infrared image ($t_{int}$ = 20 s) of the 
central 17$\arcsec \times$ 17$\arcsec$ of the Galaxy, centered on the supermassive black 
hole, Sgr A*. The image  contains $\sim$2$\times$10$^5$ stars down to $K$ = 24 mag, which 
includes $\sim$2500 known stars and a theorized population based on the observed
GC radial profile and the K-band luminosity function of the Galactic bulge.
The image includes photon, background, and read noise. 
{\em Right}: K-band luminosity functions of the planted stars ({\em red}) and
those detected ({\em blue}) by {\em StarFinder}.
}
\label{fig:gc_sim_k24}
\end{figure}

\begin{figure}[h]
\centering
\includegraphics[width=0.7\textwidth]{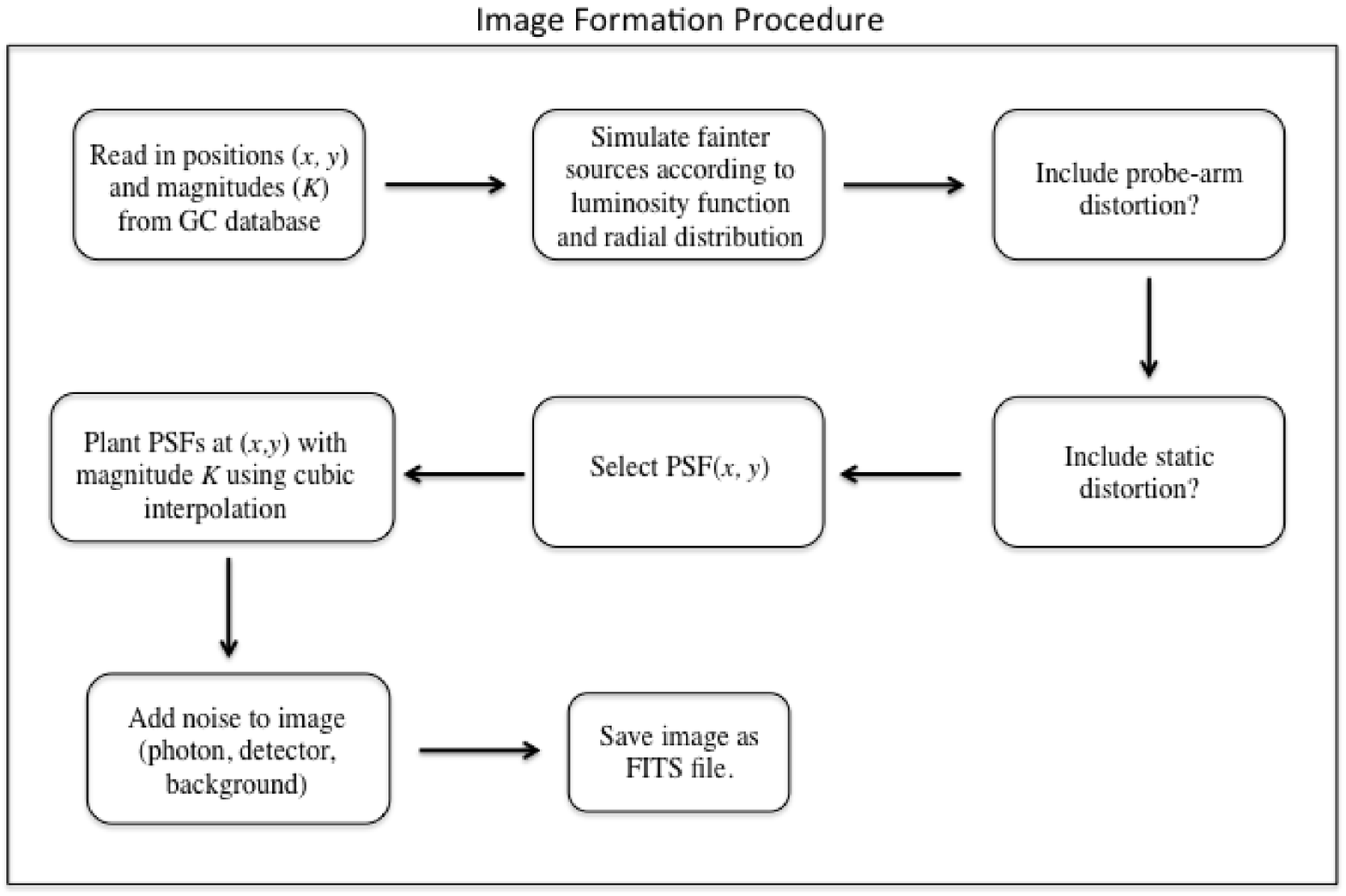}
\caption{Block diagram illustrating the necessary steps for simulating a
single image of the Galactic center.
}
\label{fig:simBlock}
\end{figure}

Galactic center simulated images include the catalog of $\sim$2500
known stars based on Keck observations ($K <$ 20) that cover a 
17$\arcsec\times$17$\arcsec$ field roughly centered on the SMBH,
Sgr A* (e.g., Yelda et al. \cite{yelda10}).  For testing the
effects of source confusion, images were also 
simulated to include fainter stars down to $K$ = 24
that will be detectable by the TMT (Figure \ref{fig:gc_sim_k24}). 
The fainter stars follow the observed radial profile of the late-type stars 
in the GC and we assume the K-band luminosity function of the 
Galactic bulge as found by Zoccali et al. \cite{zoccali03}.  
Details regarding the image simulator, including calculations of the background,
dark current, and read noise, can be found in Do et al. \cite{doIRIS}.
The general steps taken to produce a simulated image of the Galactic center are
illustrated in the block diagram in Figure \ref{fig:simBlock}.

\subsection{Astrometry}
\label{sec:ast}
Stars are identified and their positions are extracted from all images using the 
PSF fitting algorithm {\em StarFinder} \cite{stf}, which is optimized for AO 
observations of crowded stellar fields to identify and characterize stars in the 
field of view.  The user has the option of feeding in a PSF model if it is known,
or estimating the PSF from the image itself (both methods are tested here).  In the
latter case, {\em StarFinder} iteratively constructs the model based on
a set of user-selected stars in the field.  For the GC, we use a set of 25 
bright, relatively isolated stars across the field. {\em StarFinder} then 
cross-correlates the model PSF with the image in order to identify sources in the 
field. The correlation threshold is set to 0.8 for images involving only the 
known GC sources, and 0.6 for images that include the theorized fainter population.

To estimate the contribution of each error source to the error budget, the
list of input positions is compared to the list of positions extracted from the image. 
The difference between each star's positions across the two lists is computed, and
the standard deviation of these offsets ($dx$, $dy$) for stars brighter than $K$ = 17 
is reported. 

\section{Error Sources}
\label{sec:error}
\subsection{Photon \& Detector Noise}
\label{sec:baseline}
Each simulated image includes photon, detector, and thermal noise, the combination of
which may impact astrometry. To estimate this impact, we compare the extracted stellar
positions to the input positions used to simulate an individual frame (t$_{int}$ = 20 s). 
In this case, the image is created using a constant PSF over the detector, as our 
current version of {\em StarFinder} assumes a constant PSF for the astrometric analysis. 
Using the known PSF to extract positions, we find a distribution of offsets peaked near 
zero with a standard deviation along the $x$ and $y$ coordinates
of ($dx$, $dy$) = (0.00106, 0.00103) pix = (4.2, 4.1) $\mu$as.
Taking the average of the $x$ and $y$ offsets gives 4.2 $\mu$as, which
we attribute to the photon, detector, and thermal noise in the images. 
This base-level is removed in quadrature from all subsequent analyses that compare
the input and output positions, as these noise sources are included in all 
simulated images.

\subsection{Guide Probe Arm Positioning}
\label{sec:probe}
Errors in the IRIS On-Instrument Wave Front Sensor (OIWFS) guide probe positioning can 
introduce distortion errors in
astrometric measurements.  Here we test this effect by incorporating the predicted
error into the positions of Galactic center stars.

The three relatively bright and isolated guide stars that were selected have the
following positions relative to Sgra A*:
($x_1$, $y_1$) = (2.063$\arcsec$, -29.792$\arcsec$),
($x_2$, $y_2$) = (-16.857$\arcsec$, 15.379$\arcsec$), and
($x_3$, $y_3$) = (35.000$\arcsec$, 16.000$\arcsec$).
We assume that the RMS error in the probe arm positioning is 2 mas, and for each 
guide star we randomly sample an error from a Gaussian 
distribution centered on 0 mas with a 1$\sigma$ spread of 2 mas.
The positions in the catalog of the known stars in the GC were then 
distorted by applying the quadratic distortion model described in 
Sch{\"o}ck \cite{schoeck12}.

As a simple test of our ability to retrieve the original (undistorted) positions,
the distorted positions were transformed to a reference frame defined by the 
original catalog of positions, allowing for translation, rotation, and a plate
scale (independent in X and Y).
The offsets between the transformed and the original
positions have a standard deviation of 5 $\times$ 10$^{-5}$ pixels in each coordinate 
(0.2 $\mu$as). As the precision of our star lists is given to the 10$^{-5}$ pixel-level, 
this is likely an upper limit. We therefore conclude that the effect of the probe arm 
positioning errors can be entirely corrected for using a 1st-order (6-term) 
transformation.

\subsection{Static Distortion}
\label{sec:static}
\begin{figure}[h!]
\centering
\includegraphics[width=0.5\textwidth]{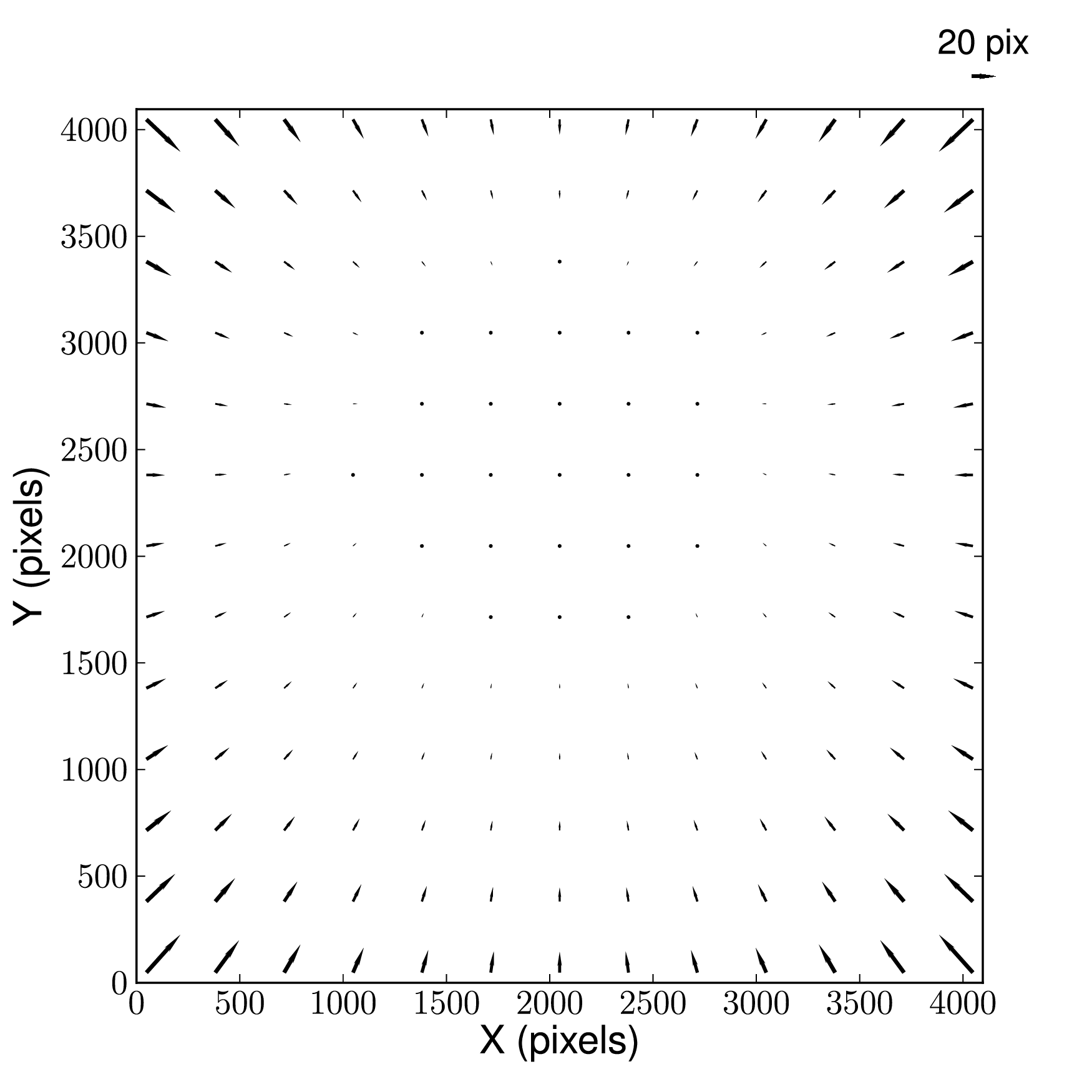}
\caption{Predicted optical distortion for the IRIS imager based on Zemax modeling
of the optics.  A 13 $\times$ 13
grid representing the positions of the pinhole mask ({\em arrow tail}) and
the expected distortion ({\em arrow head}) at those locations is shown.
}
\label{fig:dist}
\end{figure}

The static optical distortion of the IRIS imager was estimated with Zemax modeling
and is shown in Figure \ref{fig:dist}.  The distortion is shown for a grid
of 13 $\times$ 13 positions on the detector, which is the planned configuration for
the pinhole mask. We estimate that the pinholes will be drilled with a precision of
better than $\sim$15 $\mu$as\footnote{The precision with which the pinholes will be
drilled may, in reality, be larger than the 15 $\mu$as assumed here. However, using 
a self-calibration method similar to that used by Anderson \& King \cite{anderson03}, 
the distortion is expected to be known to better than this value.} and it is
assumed that this is a random Gaussian error independent from pinhole to pinhole.

The pinhole mask limits distortion measurements to a grid of positions spaced every 
$\sim$315 pixels and we must therefore interpolate between these locations in
order to find the best representation of the distortion over the entire detector.
We use a 5th order bivariate B-spline using the SciPy package {\em interpolate}.
Tests of this method reveal an interpolation error of 4.2 $\mu$as.

\begin{figure}[h!]
\centering
\includegraphics[width=0.45\textwidth]{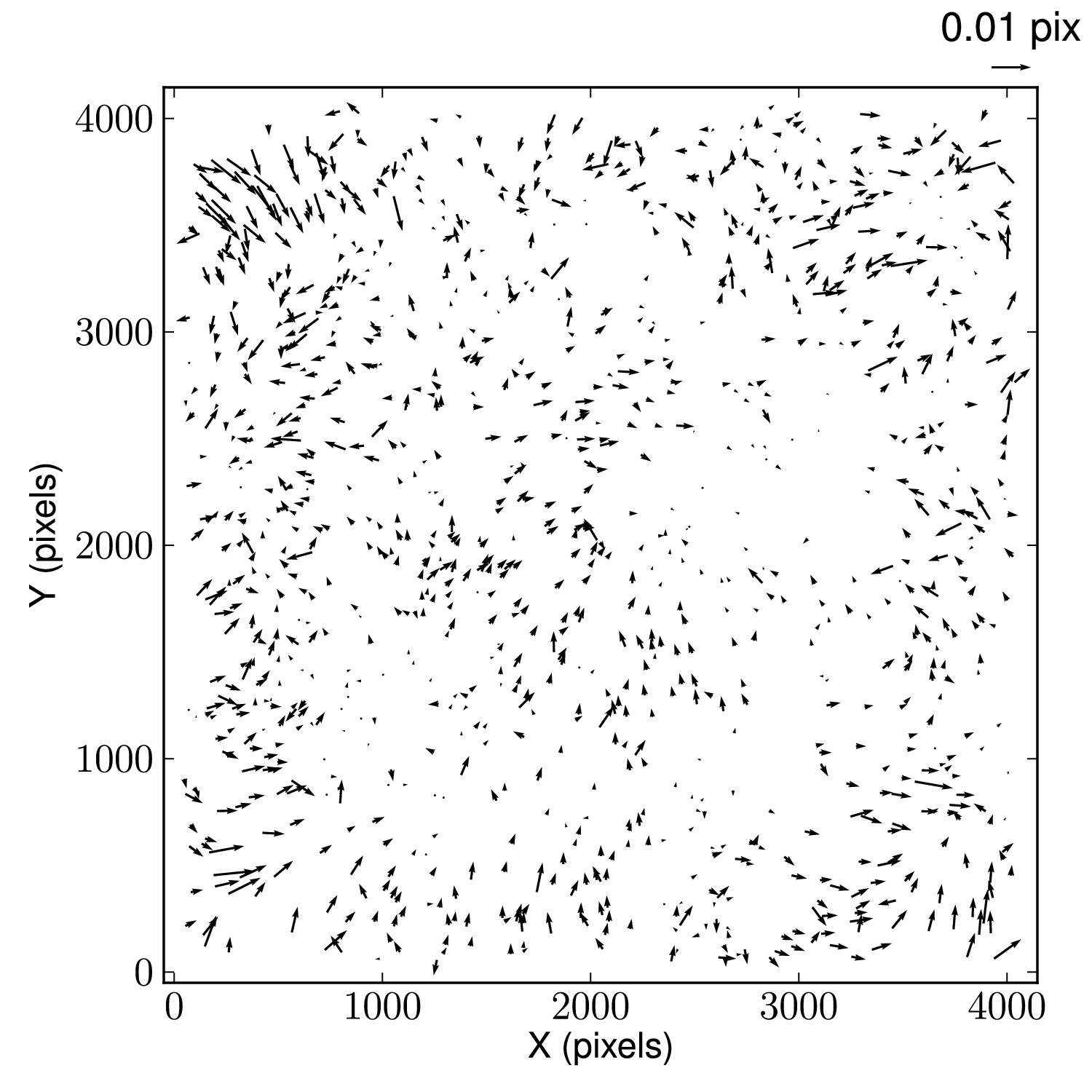}
\includegraphics[width=0.45\textwidth]{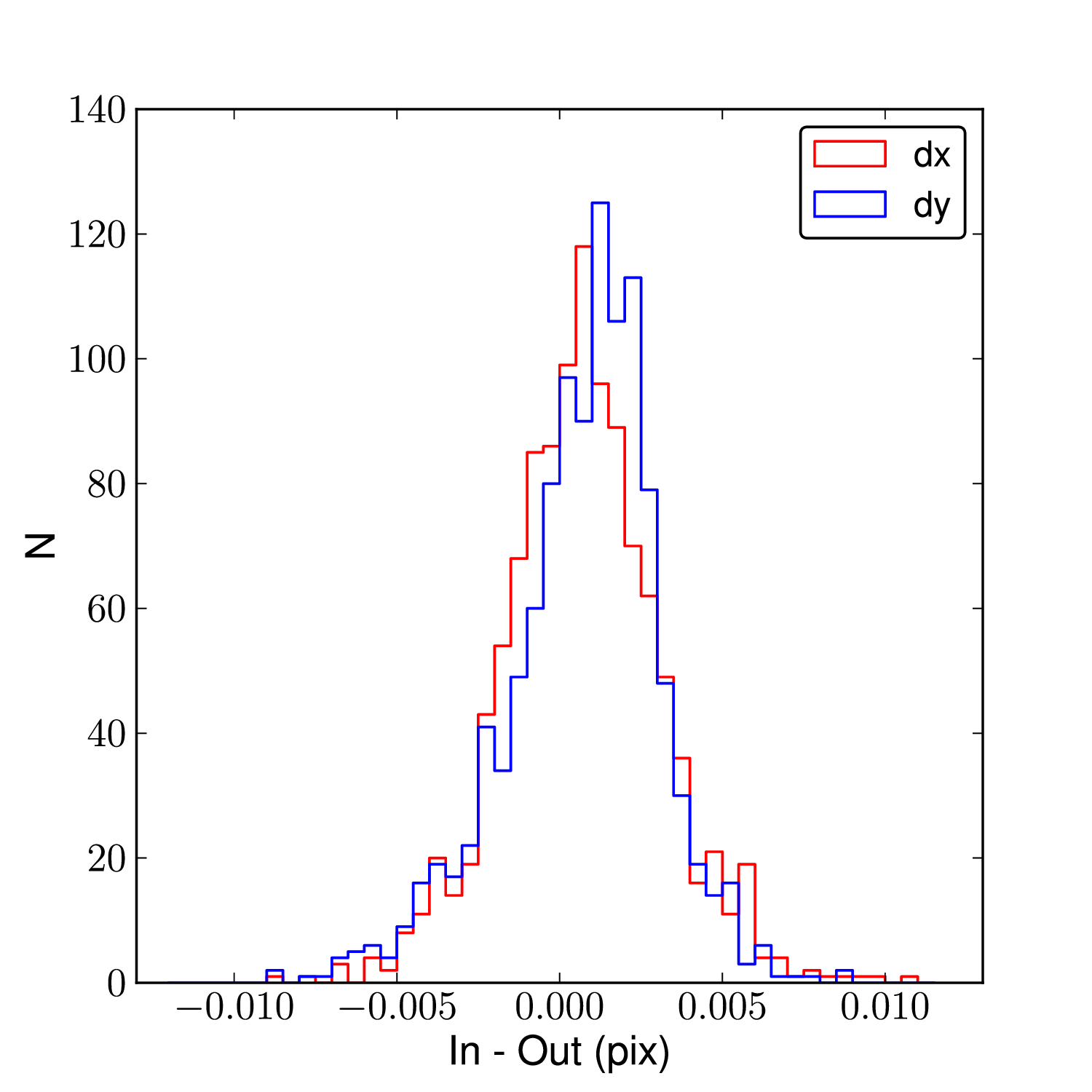}
\caption{Measured residuals after applying the distortion correction to distorted
GC stellar positions extracted by {\em StarFinder}. Note that in this case, only
the static optical distortion in IRIS is included and corrected for. The residuals 
have a 1$\sigma$ width of $\sim$0.00238 pix $\sim$ 9.5 $\mu$as. After accounting
for photon, detector, and thermal noise, as well as interpolation error, we find
the error due to residual distortion to be 7.4 $\mu$as.
}
\label{fig:GCdist}
\end{figure}

Optical distortion will be taken into account in Galactic center astrometry by 
correcting the stellar positions measured from a distorted image. We therefore
simulate such an image by distorting the positions in the catalog of known 
stars in the GC using the interpolation method described above. 
For simplicity, a spatially constant PSF is used to plant the stars.  
The positions are 
extracted from the image by {\em StarFinder} using the same PSF model and
then corrected for distortion. The difference between the original catalog of 
positions and these corrected positions is shown in 
Figure \ref{fig:GCdist}.  The distributions of residuals have a 1$\sigma$ width of
($\sigma_x$, $\sigma_y$) = (0.00238, 0.00238) pix = (9.5, 9.5) $\mu$as.
After subtracting in quadrature the error associated with photon, detector,
and thermal noise (4.2 $\mu$as), 
we find that the error due to residual distortion is 8.5 $\mu$as. 

An additional analysis was carried out in which both the probe arm 
and static distortion were included in the simulated image. The stellar positions
were again retrieved at the 8.5 $\mu$as level.

\subsection{Source Confusion}
\label{sec:confusion}
The TMT will achieve much higher sensitivities than current 8-10 m class telescopes,
allowing for the detection of stars up to $\sim$3-4 magnitudes fainter than what is
seen today.  If we extrapolate from the K-band luminosity function of the 
Galactic bulge (Zoccali et al. \cite{zoccali03}), we find that the increased 
sensitivity of TMT will result in the detection of two orders of magnitude more
stars in the central parsec of the Galaxy (i.e., from $\sim$2500 to $\sim$200,000
stars).  The presence of these additional stars can lead to source confusion, which 
in turn can introduce significant biases in astrometric measurements.

\begin{figure}[h!]
\centering
\includegraphics[width=0.5\textwidth]{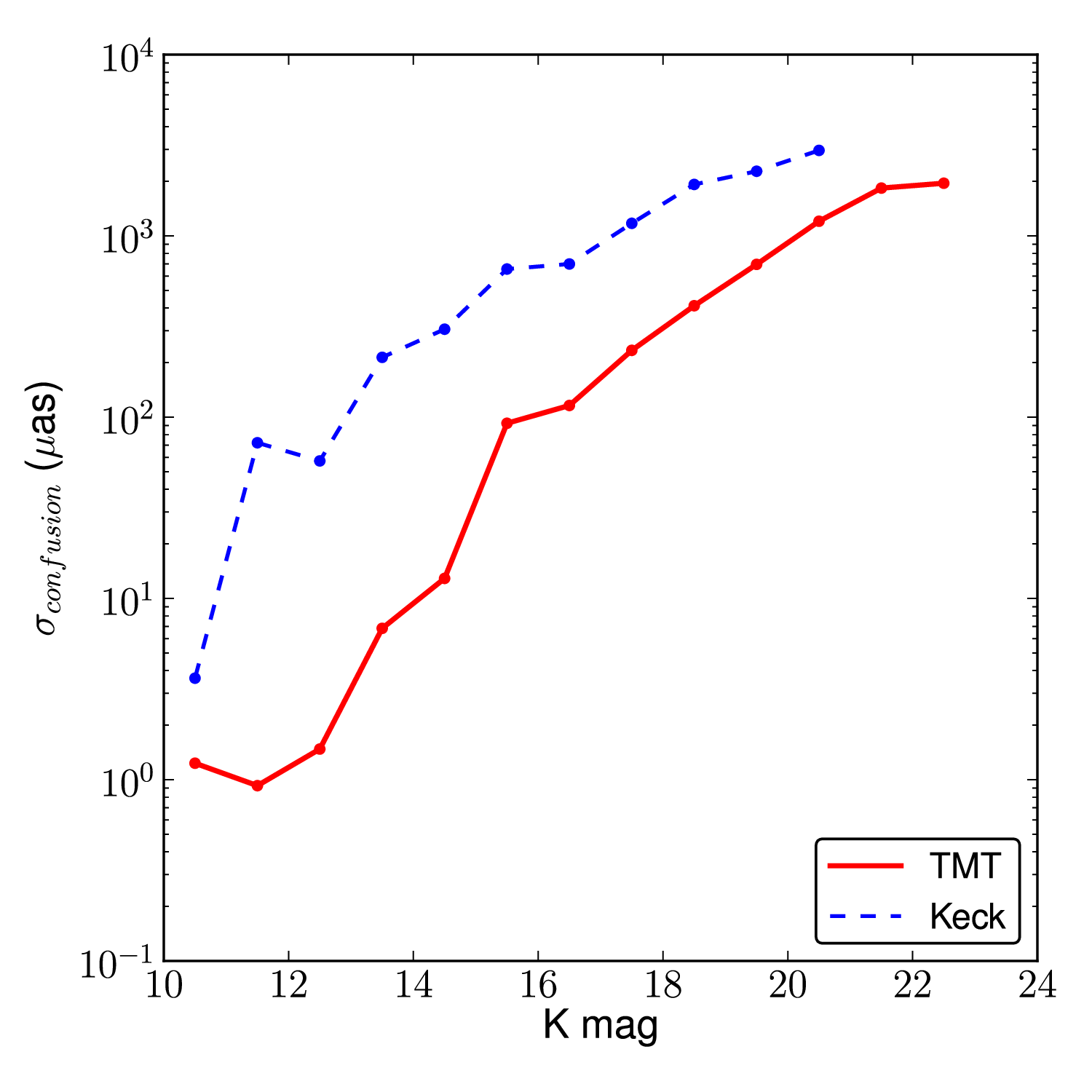}
\caption{Astrometric error resulting from confusion plotted as a function of
K-band magnitude of the planted star ({\em red solid}). Data were binned in 
1-mag bins and the RMS error of the $|$input - output$|$ positions were 
calculated and taken as $\sigma_{confusion}$. Confusion effects in Keck data are
shown for comparison ({\em blue dashed}).
}
\label{fig:confusion}
\end{figure}

We examine the effects of confusion (from both detected and undetected sources) 
using a simulated image that includes the
known GC stars as well as the theorized, fainter population (Figure 
\ref{fig:gc_sim_k24}, right). While $\sim$200,000 stars were planted, $\sim$40\% of 
the stars were detected by {\em StarFinder} at the 0.6 correlation 
threshold\footnote{Lower correlations were tested and resulted in the detection of 
a large number of spurious sources (see Yelda et al. \cite{yelda10}).}.
The extracted positions were compared to the
locations in which the stars were planted. 
Figure \ref{fig:confusion} shows the results, where we take the RMS error of the
$|$input - output$|$ positions within a magnitude bin of $dm$ = 1.
The effects of confusion for the Keck telescope are also plotted for comparison.
While confusion introduces a smaller error for TMT astrometry as compared to that
of the Keck telescope, it does become a significant term 
($\sigma_{confusion} >$ 10 $\mu$as) in the TMT error budget
for relatively faint stars ($K > $15).  For brighter stars, source confusion will
remain below $\sim$10 $\mu$as and therefore will not be the dominant source of error.

\section{Conclusions}
\label{sec:concl}
Table \ref{tab:budget} shows a breakdown of the sources of error that have been 
investigated in this report and an estimate of each source's contribution to the 
astrometric error budget.  The dominant source of error in single-epoch 
observations of the Galactic center is optical distortion 
($\sigma_{dist} \sim$ 10 $\mu$as) for bright stars ($K <$ 15) and source confusion
(($\sigma_{confusion} >$ 10 $\mu$as) for faint stars ($K >$ 15).
Future simulations will include multi-epoch observations in order
to infer proper motions and accelerations of the stars in the GC, as well as the
orbits of the closest sources to the black hole.
It will also be interesting to analyze the precision with which post-Newtonian
effects can be measured for the short-period stars.

We thank the Thirty Meter Telescope for supporting this work. We would also like to
thank Matthias Sch{\"o}ck, Brent Ellerbroek, and the rest of the TMT Astrometry
Working Group and Advisory Group for their helpful suggestions on the simulations
presented here.

\begin{deluxetable}{lc}
\tabletypesize{\scriptsize}
\tablewidth{0pt}
\tablecaption{Astrometric Error Budget for Galactic Center\label{tab:budget}}
\tablehead{
  \colhead{Effect} &
  \colhead{Error ($\mu$as)}
}
\startdata
Photon, detector, thermal noise ($t_{int}$=20 s)   	    & 4.2 \\
PSF estimation error in crowded field w/o anisoplanatism      & 3.9 \\
Guide probe positioning  	   	    & $<$0.2 \\
Static distortion			    & 8.5  \\
$\;\;\;\;$(Grid interpolation: 4.2 $\mu$as) & ~\\
$\;\;\;\;$(Distortion modeling error: 5.7 $\mu$as) & ~\\
Confusion                                   & ~\\
$\;\;\;\;$ K $<$ 15: & 1-10  \\
$\;\;\;\;$ K $>$ 15: & $>$10  \\
\enddata
\tablenotetext{a}{Static distortion assumed to be known to 15 $\mu$as.}
\end{deluxetable}

\end{document}